\documentclass{aa}
\usepackage{graphicx}
\usepackage{txfonts}
\usepackage{floatflt}
\begin{document}
  \title{Observations of V838 Mon in the CO rotational transitions}

  \author{T. Kami\'{n}ski \inst{1}, M. Miller \inst{2}, R. Tylenda \inst{1}}

  \offprints{T. Kami\'{n}ski}

  \institute{Department for Astrophysics, N. Copernicus Astronomical Centre,
            Rabia\'{n}ska 8, 87-100 Toru\'{n}, Poland\\
            \email{tomkam@ncac.torun.pl}
       \and
           I. Physikalisches Institut, Universit\"{a}t zu K\"{o}ln,
           Z\"{u}lpicher Strasse 77, 50937 K\"{o}ln, Germany}

  \date{Received; accepted}

\abstract{}{We investigate the structure of a field around the position of
V838~Mon as seen in the lowest CO rotational transitions.
We also measure and analyse emission in the same lines at
the position of V838~Mon.}{Observations have
primarily been done in the $^{12}$CO $J = 2$$\to$$1$ and $J=3$$\to$$2$ lines using
the KOSMA telescope. A field of 3.4 squared degrees has been mapped in the
on-the-fly mode in these transitions. Longer integration spectra in the
on-off mode have been obtained to study the emission at the position of
V838~Mon.
Selected positions in the field have also been observed in
the  $^{12}$CO $J = 1$$\to$$0$ transition using the Delingha telescope.}{In the
observed field we have identified many molecular clouds. They can be divided
into two groups from the point of view of their observed radial velocities.
One, having $V_{\rm LSR}$ in the range $18-32\,{\rm km\,s}^{-1}$, can be
identified with the Perseus Galactic arm. The other one, having 
$V_{\rm LSR}$ between $44-57\,{\rm km\,s}^{-1}$, probably belongs to the
Norma-Cygnus arm. The radial velocity of V838~Mon is within the second
range but the object does not seem to be related to any of the observed
clouds. We did not find any molecular buble of a $1\degr$
dimension around the position of V838~Mon claimed in van~Loon et~al.
An emission has been detected at the position of the
object in the $^{12}{\rm CO}\ J=2$$\to$$1$ and $J=3$$\to$$2$ transitions. The
emission is very narrow (FWHM~$\simeq$~1.2~km\,s$^{-1}$)
and at $V_{\rm LSR} = 53.3$~km\,s$^{-1}$. Our analysis of the data suggests 
that the emission is probably extended.}{}

        
\titlerunning{Observations of V838 Mon in CO lines.}

\authorrunning{Kami\'{n}ski et al.}

\maketitle


\section{Introduction  \label{introd_sec}}

The outburst of V838 Mon was discovered in the beginning of January~2002.
Initially thought to be a nova, the object appeared unusual and enigmatic in
its nature. The eruption, as observed in the optical, lasted about three
months (Munari et~al. \cite{muna02}, Kimeswenger et~al. \cite{kimes02},
Crause et~al. \cite{crause03}). After developing an A--F supergiant spectrum 
at the maximum at the beginning of February~2002, the object evolved to
lower effective temperatures and in April~2002 it practically disappeared
from the optical, remaining very bright in the infrared. A detailed analysis
of the evolution of the object in the outburst and decline can be found in
Tylenda (\cite{tyl05}).

Different outburst mechanisms have been proposed to explain the
eruption of V838~Mon. They include an unusual nova (Iben \& Tutukov
\cite{it92}), a late He-shell flash (Lawlor \cite{law05}) and a stellar merger
(Soker \& Tylenda \cite{soktyl03}).
These models have critically been discussed in Tylenda \& Soker
(\cite{tylsok06}) and the authors conclude that the only mechanism that
can satisfactorily account for the observational data is a collision and
merger of a low-mass pre-main-sequence star with an $\sim$$8\,M_\odot$
main-sequence star.

V838 Mon is surrounded by diffuse matter which gave rise to a spectacular
light-echo phenomenon (e.g. Bond et~al. \cite{bond03}). Bond et~al. claim
that the matter comes from previous eruptions of the object. However
Tylenda (\cite{tyl04}), Crause et~al. (\cite{crause05}), 
as well as Tylenda, Soker \& Szczerba (\cite{tss05})
argue that the echoing matter is of interstellar origin. The latter is
consistent with recent findings, namely that V838~Mon is a member of a
young open cluster (Afsar \& Bond \cite{afsar}) and that the total
mass of the echoing matter is probably of the order of $100\ M_\odot$
(Banerjee et~al. \cite{baner06}).

van Loon et al. (\cite{loon04}) have analyzed archive infrared and
radio data on the sky around V838~Mon and claimed discovery of
multiple shells ejected by 
the object in the past. In particular, from a compilation of CO galactic
surveys done in Dame et~al. (\cite{dame01}) van~Loon et~al. have suggested that
V838~Mon is situated within a bubble of CO emission with a diameter of $\sim$$
1\degr$. These results have been critically discussed in Tylenda et~al.
(\cite{tss05}), who have concluded that the shells of van~Loon et~al. are not
realistic.

Deguchi et~al. (\cite{degu05}) have discovered an SiO maser emission from
V838~Mon. The main component is narrow and centered at $V_{\rm LSR} \simeq
54\ {\rm km}\,{\rm s}^{-1}$, which is thought to be a radial velocity of
the object itself.  Further observations of Claussen et~al. (\cite{clauss06}) 
have shown that the SiO maser is variable and that most of the emission
comes from a region smaller than a milliarcsecond.

In the present paper we report on results of our observations
of V838~Mon and its nearby vicinity in the $^{12}$CO $J = 1$$\to$0, 2
$\to$1 and 3$\to$2 transitions. We describe the
observational material 
and discuss results on the CO emission from a field of 3.4 squared degrees
around the position of V838~Mon. One of the goal of this survey is to
verify the existence of the CO bubble around V838~Mon claimed in van~Loon et~al.
(\cite{loon04}).
Measurements of the CO emission obtained at the
position of V838~Mon are also presented, analysed and discussed.
A preliminary analysis of the data has been done in Kami\'{n}ski et~al.
(\cite{kmst}).

\section{Observations \label{obs_sect}}

Most of the data analysed in the present study were obtained with 
the 3~m  K\"{o}lner Observatorium f\"{u}r Sub-Millimeter Astronomie
(KOSMA) telescope in the $^{12}{\rm CO}\ J = 2 $$\to$$
1$ (230.54\,GHz) and $J=3$$\to$$2$ (345.80\,GHz) transitions. They
include maps covering an area around the position of V838~Mon as well
as several long integration measurements at 
the position of the object. Complementary observations in the $^{12}{\rm CO}\
J = 1 $$\to$$ 0$ (115.27\,GHz) transition have been obtained using the 13.7~m
Delingha telescope at the Purple Mountain Observatory (China). 
Basic technical details of the measurements are summarised in
Table~\ref{tech_tab}.

\begin{table}
\begin{minipage}[t]{\columnwidth}
\caption{Technical characteristics of the observations
  in the rotational transitions of CO.}
\smallskip
\centering
\renewcommand{\footnoterule}{} 
\begin{tabular}{lccc}
\hline
\noalign{\smallskip}
CO line& $J$ = 1$\to$0&  $J$ = 2$\to$1&  $J$ = 3$\to$2\\
telescope & Delingha & KOSMA & KOSMA \\
\noalign{\smallskip}
\hline
\noalign{\smallskip}
HPBW\footnote{half power beam width of the main beam}&55''& 130"& 82"\\
$T_{\rm sys}$ DSB\footnote{double-side band system temperature including the
  atmospheric contribution} [K]&~~230 $\div$ 290&~~150 $\div$
230&~~230 $\div$ 380\\ 
vel. range [km s$^{-1}$] &--139 $\div$ 239&--106 $\div$ 216&--121 $\div$
231\\
$\Delta V_{\rm ch}$ [km s$^{-1}$]&0.37&0.22&0.29\\
$\eta_{\rm mb}$  & 0.63 & 0.68 & 0.79 \\
$<$$\sigma_{\rm rms}$$>$\footnote{average sensitivity given in the
  main beam scale} [K] &0.14&0.42&0.93\\
\noalign{\smallskip}
\hline
\end{tabular}
\label{tech_tab}
\end{minipage}
\end{table}

\begin{figure}
 \resizebox{\hsize}{!}{\includegraphics[trim = 0 0 0 100]{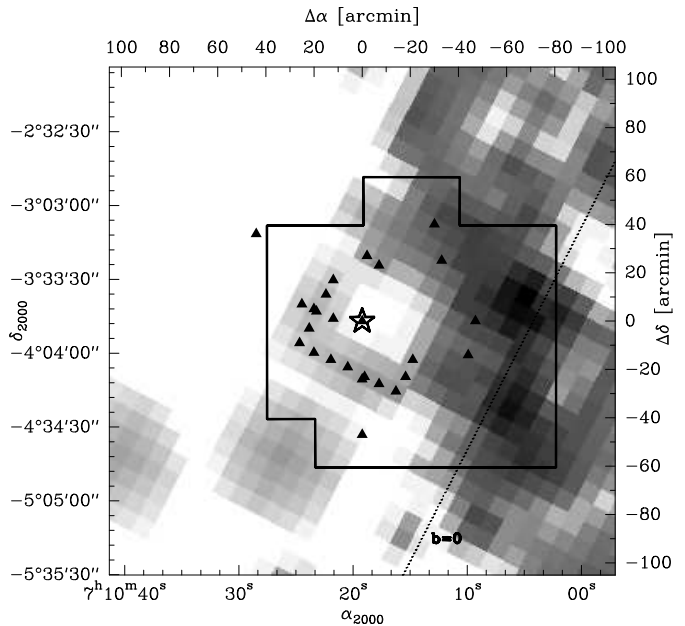}}
 \caption{A $3\fdg5 \times 3\fdg5$ map centered at
the V838~Mon position from the Dame et~al. (\cite{dame01}) compilation 
of CO (1$\to$0) surveys (the data taken from {\it SkyView} -- {\tt \small
    http://skyview.gsfc.nasa.gov/}). The star symbol marks 
the position of V838~Mon. The polygon drawn with a solid line shows the region
mapped in our KOSMA on-the-fly observations in the CO (2$\to$1) and 
(3$\to$2) transitions. Filled triangles mark
the positions observed with the Delingha telescope in the CO (1$\to$0)
transition. Dashed line indicates the Galactic equator. The upper and
right axes show the offsets in arcminutes from the position of V838~Mon.}
 \label{fig_1}
\end{figure}

\subsection{Observations with the KOSMA telescope  \label{kosma_sect}}
 
The main observations mapping an area of 3.4 squared degrees around the
position of V838~Mon were acquired on 1-5~April~2005. The maps were obtained
in the on-the-fly observing mode (Kramer et al. \cite{kram99}, Beuther et
al. \cite{beut00}) with a sampling of $1\arcmin$ and with an emission-free
reference position at $\alpha_{\rm 2000}=07^h\,07^m\,00^s$, $\delta_{\rm 2000}=
-03\degr\,15\arcmin\,00\arcsec$. Observations resulted in a set of
12\,400 spectra in each of the two transitions (2$\to$1 and 3$\to$2). Typical integration time was of 4~s.  

Figure~\ref{fig_1} shows a $3\fdg5 \times 3\fdg5$ map centered at
the V838~Mon position taken from the Dame et~al. (\cite{dame01}) compilation 
of CO (1$\to$0) surveys. The region mapped in our observations is shown
with a solid line.  

A number of long integration spectra in 
both transitions at the position of V838~Mon were
obtained on 4~April~2005 and 26~December~2005. On 16-19 April~2006
observations were done only in the (2$\to$1) line, but apart from the
object position, 6 nearby positions with offsets ($0,1$), ($0,2$),
($0,-1$), ($0,-2$), ($-2,0$), and ($2,0$) ($\Delta\alpha$, $\Delta\delta$
-- in arcmin, relative to the V838 Mon position) were also observed.

During the observations the KOSMA telescope (Kramer et
al. \cite{kram98}) was equipped with a 
dual-channel SIS receiver (Graf et al. \cite{graf98}) with a bandwidth
of 1~GHz. The receiver was connected to acousto-optical 
spectrometers of medium and variable resolution
(Schieder et al. \cite{schieder89}) with a channel spacing (bandwidth)
of 165.4~kHz (248~MHz) and 339~kHz (407~MHz) for the 
CO (2$\to$1) and (3$\to$2) data sets, respectively. The system temperature
was changing significantly during the observations mainly due to
variable atmospheric conditions. It was in a range of 150--230~K at
230~GHz and 230--380~K at 345~GHz. Pointing scans on
planets were carried out every 2--3~hours and the pointing accuracy
was better than about 10$\arcsec$. The data were calibrated using the standard
chopper wheel method (Kutner \& Ulich \cite{kutner81}) giving spectra
in the antenna temperature, $T_{\rm A}^\ast$ (e.g. Rohlfs \& Wilson
\cite{rohlfs04}), corrected for atmospheric attenuation, ohmic losses, rearward
spillover and scattering. In order to remove significant instrumental baseline
effects, polynomial baselines up to fourth order were subtracted.  

\subsection{Observations with the Delingha telescope  \label{delingha_sect}}

Observations in the $^{12}$CO $J=1$$\to$$0$ transition were conducted
on 28~October -- 3~November~2005. 
25 points around the object were measured. All the points are
shown with triangles in Fig.~\ref{fig_1}. As can be seen from
Fig.~\ref{fig_1}, most of the points are located in the bubble-like
structure claimed in van~Loon et~al. (\cite{loon04}). 
The position of V838~Mon as well as 8 nearby positions with
offsets ($0,1$), ($-1,1$), ($-1,0$), ($-1,-1$), ($0,-1$), ($1,-1$), ($1,0$),
and ($1,1$) ($\Delta\alpha$, $\Delta\delta$ --
in arcmin, relative to the V838 Mon position) have also been measured.
The integration time was 5 minutes for the object position and 1 minute per each
of the 8 offset positions.

All the Delingha observations were done in the position
switching mode with a reference position at $\alpha_{\rm
  2000}=07^h\,07^m\,01^s$, $\delta_{\rm 2000}=-
03\degr\,14\arcmin\,51\arcsec$. Each point was observed with an
integration time of $4-8$~min. A heterodyne SIS receiver and an acousto-optical
spectrograph have been used as frontend and backend,
respectively. A typical system temperature was of 260~K~(DSB,
including sky contribution). Bandwidth and channel spacing were
145.5~MHz and 142~kHz,
respectively. Pointing of the antenna was regularly checked by
observing planets and SiO masers at 86~GHz and had an accuracy of
6$\arcsec$. The observations were calibrated and expressed in the
antenna temperature scale. From the resulting spectra low order polynomial
baselines were subtracted. \\

Most of the results presented in the next section are in the scale of the
main beam temperature, $T_{\rm mb}$. This is related to the antenna
temperature as $T_{\rm mb} = T_{\rm A}^\ast/\eta_{\rm mb}$, where $\eta_{\rm
mb}$ is the main beam efficiency given for each transition in
Table~\ref{tech_tab}. Beamwidths of the antennas at the appropriate
frequencies are also presented in the table. The last raw in
Table~\ref{tech_tab} gives the sensitivity of our observations in
terms of an average baseline-noise rms of the spectra. All velocities
in this paper are given with respect to the local standard of rest (LSR).

All the data reduction as well as the analysis of the
results have been done using the GILDAS\footnote{{\it Grenoble Image and Line
Data Analysis Software}, {\tt www.iram.fr/IRAMFR/GILDAS}} software. 

\section{Maps of the field around the V838~Mon position  \label{maps_sect}}

The observations with the KOSMA telescope provided us with a total number of 24\,800
spectra in the $J = 2$$\to$$1$ and 3$\to$2 transitions. The resultant maps in
both transitions are qualitatively similar. The survey in $J = 2$$\to$$1$
was however more sensitive (see Table~\ref{tech_tab}) and gave more details in the
maps, so most of the discussion and estimates done in the present
section will be based on the $J = 2$$\to$$1$ survey. 

\begin{figure}
 \resizebox{\hsize}{!}{\includegraphics[angle=270]{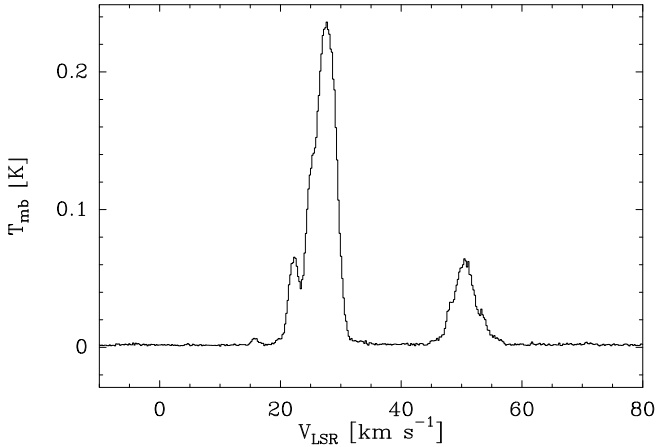}}
 \caption{The mean spectrum obtained from averaging over all the observed
positions in the $J=$~2$\to$1 transition.}
 \label{mean_sp}
\end{figure}

Figure \ref{mean_sp} shows a mean spectrum obtained from averaging over all
the observed positions in the $J=2$$\to$$1$ transition. As can bee seen,
a significant signal has been detected in two well separated ranges of
$V_{\rm LSR}$, namely $18-32$ and $44-57\ {\rm km}\,{\rm s}^{-1}$. 
As discussed in Appendix~\ref{append}, they
correspond to two populations of interstellar clouds belonging to two
different spiral arms (Perseus arm and Norma-Cygnus arm, respectively) of the Galaxy.
The second velocity range ($44-57\ {\rm km}\,{\rm s}^{-1}$) is more interesting
from the point of view of V838~Mon. The presumable radial velocity of the
object, as inferred e.g. from the SiO maser emission, is within this range.

\begin{figure}
 \resizebox{\hsize}{!}{\includegraphics{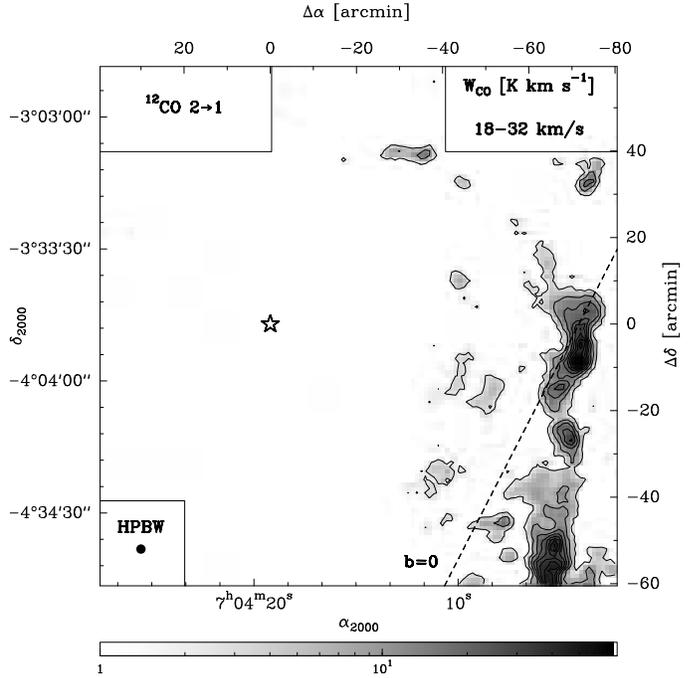}}
 \caption{The intensity map of CO $J$ = 2$\to$1 integrated over a
  the $V_{\rm LSR}$ range 18--32~km~s$^{-1}$. Contours are plotted from
  2.7 to 63.8~K~km~s$^{-1}$ with a spacing of 6.8~K~km~s$^{-1}$ 
  (4~to~94\% spaced with 10\% of the maximum). 
  The star-like symbol indicates the V838~Mon position. The
  Galactic equator is shown as a dashed line.} 
 \label{map_1}
\end{figure}

\begin{figure}
 \resizebox{\hsize}{!}{\includegraphics{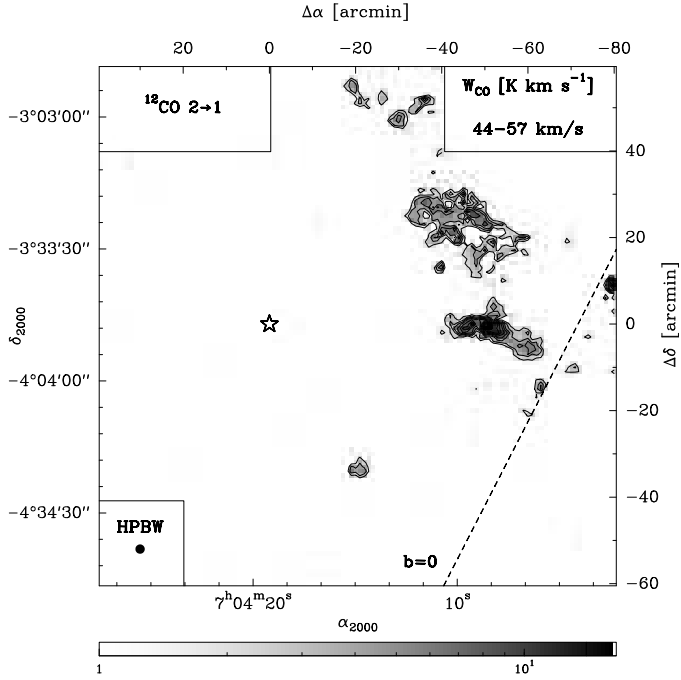}}
 \caption{The same as Fig.~\ref{map_1} but integrated over the
  $V_{\rm LSR}$ range 44--57~km~s$^{-1}$. 
  Contours are plotted from 2.3 to 15.5~K~km~s$^{-1}$ with a spacing of
  1.6~K~km~s$^{-1}$ (14~to~94\% spaced with 10\% of the maximum).}
 \label{map_2}
\end{figure}

Figure \ref{map_1} presents a map of the $J=2$$\to$$1$ emission integrated over
the $V_{\rm LSR}$ range $18-32$~km~s$^{-1}$. The
same but for the $V_{\rm LSR}$ range $44-57$~km~s$^{-1}$, is shown in Fig.~\ref{map_2}. Channel maps derived from  
integrations over narrower velocity ranges, both in the $J=2$$\to$$1$ and
3$\to$2 transitions, as well as a detailed discussion of the molecular
clouds identified from the maps, can be found in Appendix~\ref{append}.


\subsection{On the bubble-like structure of van Loon et al. (\cite{loon04})
\label{buble_sec}}

As mentioned in Sect.~\ref{introd_sec} one of our motivations to observe the
surroundings of V838~Mon in the CO rotational lines
was the claim of van~Loon et~al. (\cite{loon04}) that the object is situated
in a bubble-like structure of $\sim$$1\degr$ in diameter seen in the
CO~(1$\to$0) transition. As can be easily seen from our maps in 
the (2$\to$1) line no structure of this kind is present (see Figs.
\ref{map_1} and \ref{map_2}). In particular, if the structure were real and
related to V838~Mon it should have been present in the $V_{\rm LSR}$
range presented in in Fig.~\ref{map_2}, which is not the case.

Most of our observations obtained with the
Delingha telescope were meant to further investigate this problem. These
measurements were done in the CO~(1$\to$0) transition, i.e. in the same line
in which the map investigated by van~Loon et~al. had been done, 
and the observed positions
were chosen to probe the structure of van~Loon et~al. (see
Fig.~\ref{fig_1}). Note also that the Delingha observations were significantly
more sensitive than the KOSMA maps (see Table~\ref{tech_tab}).

A significant emission at a level of 4$-$8~${\rm K\,km\,s}^{-1}$ was detected
at 4 positions with the following ($\alpha, \delta$) offsets (in arcmin) relative 
to the V838~Mon position: $(-30, 40),\ (-33, 25),\ (-44, -14)$, and $(-47, 0)$
(4 rightmost triangles in Fig.~\ref{fig_1}). These detections are within
the emission regions seen in our (2$\to$1) maps obtained with KOSMA
(Figs.~\ref{map_1} and \ref{map_2}). A less significant emission at
a level of $0.6-1.5\ {\rm K\,km\,s}^{-1}$ was also found at $(-7, 23),\ (12, 1)$,
and $(6, -19)$. No emission at the latter positions was detected in the KOSMA survey,
most probably due to a lower sensitivity of the latter. No emission in the
(1$\to$0) transition was detected at any other of the remaining 18 positions 
observed with the Delingha telescope. 

We can thus conclude that our data do not confirm the existence of 
the CO buble-like structure claimed in
van~Loon et~al. (\cite{loon04}). Note that this structure
was not seen in any other domain, in particular in the IRAS maps (e.g.
Tylenda et~al. \cite{tss05}). As discussed in Tylenda et~al. (\cite{tss05}), 
the CO bubble of van~Loon et~al. was likely to be an
artefact of merging two surveys of different resolution and sensitivity
in the compilation of Dame et~al. (\cite{dame01}).

\section{Observations at the position of V838 Mon  \label{obs_object}}

\subsection{Results from the KOSMA telescope  \label{kosma}}

\begin{figure}
 \resizebox{\hsize}{!}{\includegraphics{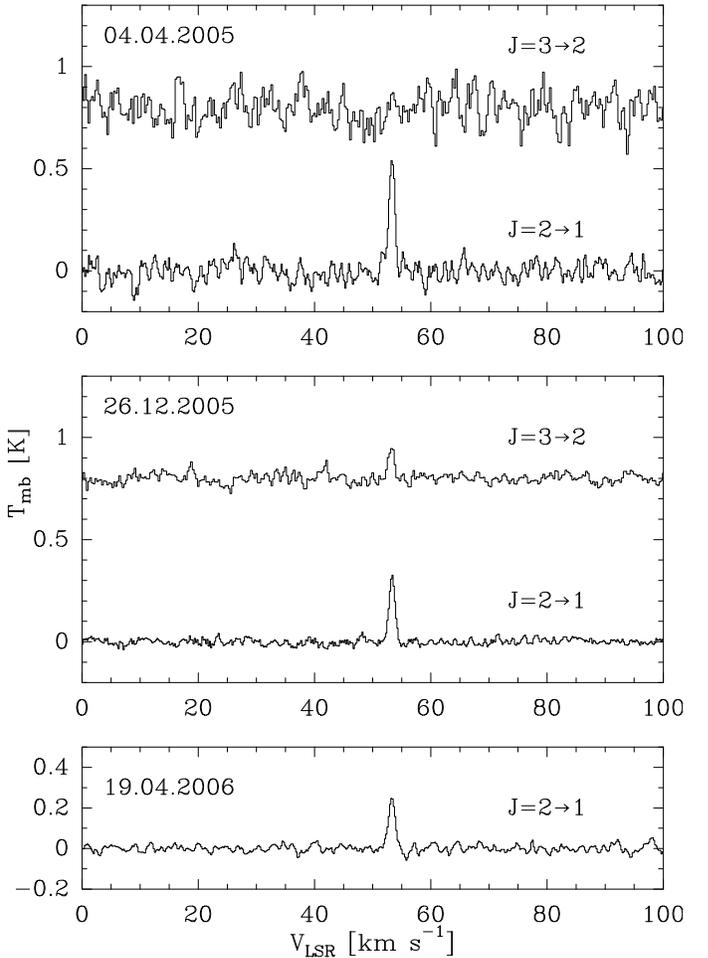}}
 \caption{Spectra in the CO (2$\to$1) and (3$\to$2)
transitions obtained with the KOSMA telescope at the position of V838~Mon.}
 \label{spect_fig}
\end{figure}

The on-the-fly observations of the field reported in Sect.~\ref{maps_sect}
and Appendix~\ref{append} did not
reveal any significant signal at the position of V838~Mon. These were
however measurements mostly done with an integration time of 4 sec. Longer
observations in a standard on-off mode, 
reported in Table~\ref{res_tab} and displayed in
Fig.~\ref{spect_fig}, have however shown a clear emission. The line is very
narrow and centered at a velocity very close to that of the SiO maser 
($V_{\rm LSR} = 54.1\ {\rm km\,s}^{-1}$, Deguchi et~al. \cite{degu05}). In
April~2005 we did not detect any emission in the (3$\to$2) transition and an
upper limit to $T_{\rm mb}$ of 0.25~K ($3\sigma$) can be set up. More sensitive
observations in December~2005 confirmed that the emission is also present in
the (3$\to$2) transition at the same velocity as the (2$\to$1)
component but about twice fainter. 

\begin{table*}
\begin{minipage}[t]{16.4cm}
\caption{Results of the CO observations at the V838 Mon position with KOSMA.}
\label{res_tab}
\smallskip
\renewcommand{\footnoterule}{} 
\begin{tabular}{ccrccccccc}
\hline
\noalign{\smallskip}
date&line&HPBW
&$T_{\rm sys}$&$\Delta V_{ch}$&$\sigma_{\rm rms}$
&\multicolumn{1}{c}{$V_{\rm
    LSR}$}&\multicolumn{1}{c}{Peak $T_{\rm mb}$}&FWHM&$I_{\rm
  CO}$\footnote{integrated intensity $I_{\rm CO}=\int T_{\rm mb}(V) \;dV$}\\ 
&&[arcsec]&[K]&[km~s$^{-1}$]&[K]&[km~s$^{-1}$]&\multicolumn{1}{c}{[K]}&[km~s$^{-1}$]&[K
  km~s$^{-1}$]\\ 
\noalign{\smallskip}
\hline
\noalign{\smallskip}
04.04.2005 & 2--1 & 130 & 168 & 0.21 & 0.043 & 53.3 & 0.553 & 1.17 & 0.691\\
           & 3--2 &  82 & 273 & 0.29 & 0.083 & ...  & ...   & ...   & ... \\
26.12.2005 & 2--1 & 130 & 310 & 0.21 & 0.014 & 53.3 & 0.323 & 1.18 & 0.406\\
           & 3--2 &  82 & 418 & 0.29 & 0.022 & 53.2 & 0.156 & 1.50 & 0.190\\
19.04.2006 & 2--1 & 130 & 397 & 0.21 & 0.017 & 53.3 & 0.249 & 1.34 & 0.354\\
\hline
\noalign{\smallskip}
\end{tabular}
\end{minipage}
\end{table*}

\begin{figure}
 \resizebox{\hsize}{!}{\includegraphics[angle=270]{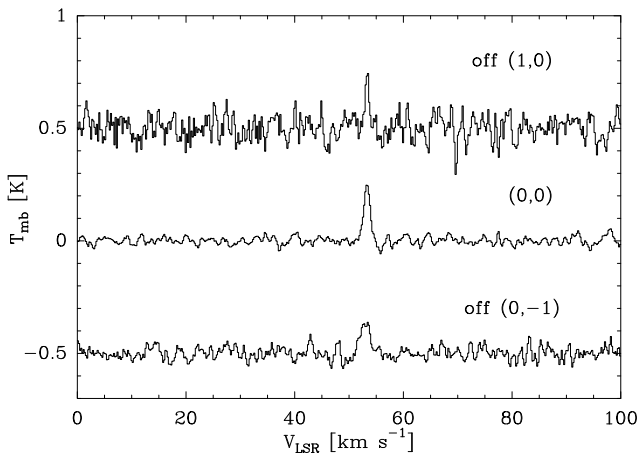}}
 \caption{Spectra in the CO (2$\to$1)
transition obtained on 16-19~April~2006 at the position of V838~Mon (middle)
and at two offsets (in $\alpha$ and $\delta$ in arcmin).}
 \label{spect2_fig}
\end{figure}

Observations done in April~2006 reavealed a significant emission not only at
the position of V838~Mon but also at two offset positions, namely at ($0,1$)
and ($0,-1$). The obtained spectra are shown in Fig.~\ref{spect2_fig}, while
the results of measurements can be found in Table~\ref{res2_tab}. No
significant emission has been detected in any of the other 4 offset
positions and an upper limit to  $T_{\rm mb}$ is 0.11~K ($3\sigma$).

\begin{table}
\caption{Results of the CO (2$\to$1) observations on 16-19 April 2006.}
\label{res2_tab}
\begin{tabular}{cccccc}
\hline
\noalign{\smallskip}
offset&$\sigma_{\rm rms}$&$V_{\rm LSR}$&Peak $T_{\rm mb}$&FWHM&$I_{\rm CO}$\\
\,[arcmin]&[K]&[km~s$^{-1}$]&[K]&[km~s$^{-1}$]&[K km~s$^{-1}$]\\
\noalign{\smallskip}
\hline
\noalign{\smallskip}
($0,1$) & 0.052 & 53.3 & 0.251 & 0.82 & 0.220\\
($0,0$) & 0.017 & 53.3 & 0.249 & 1.34 & 0.354\\
($0,-1$)& 0.027 & 53.0 & 0.139 & 2.00 & 0.297\\
\hline
\noalign{\smallskip}
\end{tabular}
\end{table}

\subsection{Results from the Delingha telescope  \label{delingha}}

No emission has been detected at the V838~Mon position in the
(1$\to$0) transition and an upper limit 
to $T_{\rm mb}$ is 0.39~K ($3\sigma$).
Weak emission has been detected at three offsets and the results of
measurements are in Table~\ref{resd_tab}. The features are, however, at a
level of $\sim$$2\sigma$, so one can doubt if they are real. The feature at
(0,1) is at the same velocity as the emissions seen in the (2$\to$1)
transition, which may argue that the feature is real. On the other hand the
features in the two other offsets are at different velocities and wider than
the lines observed in the (2$\to$1) transition, which raises additional doubts
about their reality.

\begin{table}
\caption{Results of the CO (1$\to$0) observations obtained with the Delingha
telescope on 28~Oct.--3~Nov.~2005.}
\label{resd_tab}
\begin{tabular}{cccccc}
\hline
\noalign{\smallskip}
offset&$\sigma_{\rm rms}$&$V_{\rm LSR}$&Peak $T_{\rm mb}$&FWHM&$I_{\rm CO}$\\
\,[arcmin]&[K]&[km~s$^{-1}$]&\multicolumn{1}{c}{[K]}&[km~s$^{-1}$]&[K km~s$^{-1}$]\\
\noalign{\smallskip}
\hline
\noalign{\smallskip}
($0,1$) &0.39&53.5&0.83&2.79&2.46\\
($0,-1$)&0.27&48.5&0.73&3.25&2.52\\
($1,-1$)&0.35&48.2&0.53&7.49&4.20\\
\hline
\noalign{\smallskip}
\end{tabular}
\end{table}

\section{Discussion and conclusions  \label{disc}}

Our on-the-fly maps in the CO rotational transitions, especially those done in
the (2$\to$1) line with the KOSMA telescope, have allowed us to identify 29
molecular clouds around the position of V838~Mon (see Appendix~\ref{append}).
As far as we know, the region mapped with the KOSMA telescope has
not been observed before with a sensitivity
and a spatial resolution comparable to our survey. 
The data presented in  Appendix~\ref{append} can be used in studies of
molecular matter in the outer Galaxy. 

We do not confirm the existence of a molecular bubble of a 1 degree dimension
claimed in van~Loon et~al. (\cite{loon04}),
which was probably an artefact resulted
from merging two surveys of different resolution and sensitivity in the data
used by van~Loon et~al.

Our maps did not reveal any molecular cloud in the near vicinity
of V838~Mon. The nearest CO emission has been detected at $\sim$$40$~arcmin
from the position of the object (see Fig.~\ref{map_2}). Thus the nearest
dense molecular cloud is located at least at a distance of $\sim$$80$~pc 
from V838~Mon (assuming a 7~kpc distance to V838~Mon -- see below). From
the noise level in the (2$\to$1) map (see Table~\ref{tech_tab}) we can put
an upper limit to $I_{\rm CO}=\int T_{\rm mb}\,dV$ of
$1.26~{\rm K~km~s}^{-1}$ ($3 \sigma_{\rm rms}$) at the 
position of V838~Mon. Assuming the $N$(H$_2$) to $I_{\rm CO}$
conversion factor of $X_{\rm CO} = 5.1$ (in units 
$10^{20}\,{\rm molecules}\ {\rm cm}^{-2}\,{\rm K}^{-1}\,{\rm
  km}^{-1}\,{\rm s}$, typical value for the NC clouds discussed in
Appendix~\ref{append}) we obtain an upper limit to the 
column density of $N({\rm H}_2) = 6.4 \times 10^{20}~{\rm cm}^{-2}$. This
upper limit is an order of magnitude lower than typical column densities observed for 
molecular clouds in the vicinity of the Sun (e.g. Blitz \cite{blitz}).
Thus we can conclude that V838~Mon is not located in a typical molecular
cloud. 

However, as presented in Sect.~\ref{kosma}, long integrations in the on-off
mode have allowed us to detect an emission in the CO
(2$\to$1) and (3$\to$2) transitions at the position of V838~Mon. The question
that arises is: what is the origin of this emission? Certainly it cannot come
from the matter ejected in the 2002 outburst. That matter was expanding
with large velocities, a few hundred km~s$^{-1}$, while our profiles have
a FWHM of $\sim$$1\,{\rm km\,s}^{-1}$. 

The $V_{\rm LSR}$ position and width of
the CO lines similar to those of the SiO maser (Deguchi et~al.
\cite{degu05}, Claussen et~al. \cite{clauss06}) suggest that both emissions
(CO and SiO) originate from the same place,
e.g from the remnant of the 2002 outburst. However,
in this case the KOSMA telescope would see a point source, which does not
seem to be the case. The April 2006 observations show detectable emissions
at the 1~arcmin offsets from the V838~Mon position (see
Table~\ref{res2_tab}). The HPBW of the KOSMA beam at 230~GHz
is 130~arcsec. If the CO emission source were a point source at the (0,0)
position, than the intensity measured at the (0,1) and ($0,-1$) offsets 
would be about
twice fainter than that at the central position. Table~\ref{res2_tab} shows
that this ratio is larger, i.e. $0.6-0.8$. Unfortunantely the accuracy of
the measurements at the 1~arcmin offsets was not high, so we cannot say that
the results in Table~\ref{res2_tab} are definitively inconsistent with a
point source emission.

However, there are other arguments in favour of the idea that the CO emission
is extended and/or situated outside the V838~Mon position. Our Delingha
observations in the (1$\to$0) transition (see Sect.~\ref{delingha})
have not detected any emission at the
V838~Mon position and the upper limit was 0.39~K. Assuming that the
intensity in the (1$\to$0) line is comparable to that in the (2$\to$1) line
(which is usually the case in molecular clouds) and that we observe a point
source at the V838~Mon position than from the measured $T_{\rm mb}$ in the
(2$\to$1) transition in December~2005 (see Table~\ref{res_tab}), 
taking into account different beamsizes of
the two telescopes (55~arcsec in Delingha versus 130~arcsec in KOSMA), the
expected value of $T_{\rm mb}$ in the (1$\to$0) Delingha observations would
be $\sim$$1.8$\,K. This is 4.5 times higher than the observed upper limit.
Thus either the (2$\to$1)/(1$\to$0) ratio is exceptionally large ($> 4.5$) or
the CO source is situated outside the Delingha beam but still inside the
KOSMA beam. The later interpretation is supported by our possible detection
of an emission with Delingha at three positions $\sim$$1$\,arcmin from the object
(see Table~\ref{resd_tab}).
It is also supported by a finding of Deguchi et~al. (\cite{degu06})
who, using the Nobeyama telescope, tried to measure 
the CO (1$\to$0) emission from a few
points in a field around V838~Mon. A signal was detected from a position
30~arcsec north of V838~Mon at a velocity very close to that of the
(2$\to$1) line in our KOSMA observations. No emission was however recorded 
at the position of the object.

The above discussion allows us to conclude that the CO emission, clearly
seen in our KOSMA observations, most probably originates not from V838~Mon
itself but from a region (regions) situated, typically, $\sim$$1$\,arcmin from the
V838~Mon position. There are two possible ways of explaining the emission in
this case. One is that the observed emission is a faint part of a larger CO structure
belonging, for instance, to a molecular cloud. However, as discussed above, 
our CO maps have not revealed any dense CO cloud of similar $V_{\rm LSR}$ 
closer than $\sim$$40$\,arcmin from the position of V838~Mon. 

All the
detected and possibly detected CO emission at and near the position of
V838~Mon lie well within the optical light echo of V838~Mon. Hence the second
possibility, namely that the CO emission comes from the same matter that
gave rise to the light echo and that it was the light flash from the 2002
eruption which induced the emission. Then the observed narrowness of the
line profile would be a strong argument that the matter is of interstellar
origin rather than being ejected by V838~Mon in previous eruptions. The
observed $V_{\rm LSR} = 53.3\,{\rm km\,s}^{-1}$ of the CO emission would
imply a kinematical distance of $\sim$$7.0$\,kpc (using the Galactic
rotational curve of Brandt \& Blitz \cite{bb93}). This can be compared to
$6.1 - 6.2$~kpc found by Bond et~al. (\cite{bond06}) and $\sim$$8$~kpc
advocated in Tylenda (\cite{tyl05}). 

Rushton et al. (\cite{rush03}) searched for CO emission from V838~Mon 
about a year after the 2002 eruption. No measurable signal was detected 
in the three lowest rotational transitions and the upper limit was 
$T_A^* \la 25-40$~mK. It is not straightforward to compare this result with
our findings as the observations of Rushton et~al. were done about 3~years
before ours and the object probably evolved significantly during this time
span. Nevertheless, given the beamwidth of the telescopes used by Rushton
et~al. (HPBW~$\le 45\,\arcsec$), their negative result at the position of
the object does not preclude a possibility that a significant emission was 
present in a near vicinity of the object, but outside the telescope beam.

More sensitive observations with a
higher spatial resolution are required to distinguish between the above
discussed possibilities and to futher investigate the problem of the CO
emission from V838~Mon.

\begin{acknowledgements} We wish to thank all the staff at Delingha, the
millimeter-wave radio telescope of Purple Mountain Observatory (China), for
the observations in the CO $(1$$\to$$0$) transition.

The KOSMA 3 m telescope is operated by the K\"{o}lner Observatorium f\"{u}r
SubMillimeter Astronomie of the I. Physikalisches Institut, Universit\"{a}t
zu K\"{o}ln in collaboration with the Radioastronomisches Institut,
Universit\"{a}t Bonn.

The research reported in this paper was partly supported from a grant no.
N203\,004\,32/0448 financed by the Polish Ministry of Science and Higher 
Education.
\end{acknowledgements}

\Online

\begin{appendix}
\section{Channel maps and the ISM cloud parameters}\label{append}

\begin{figure*}
\centering
\includegraphics[width=16.4cm,clip]{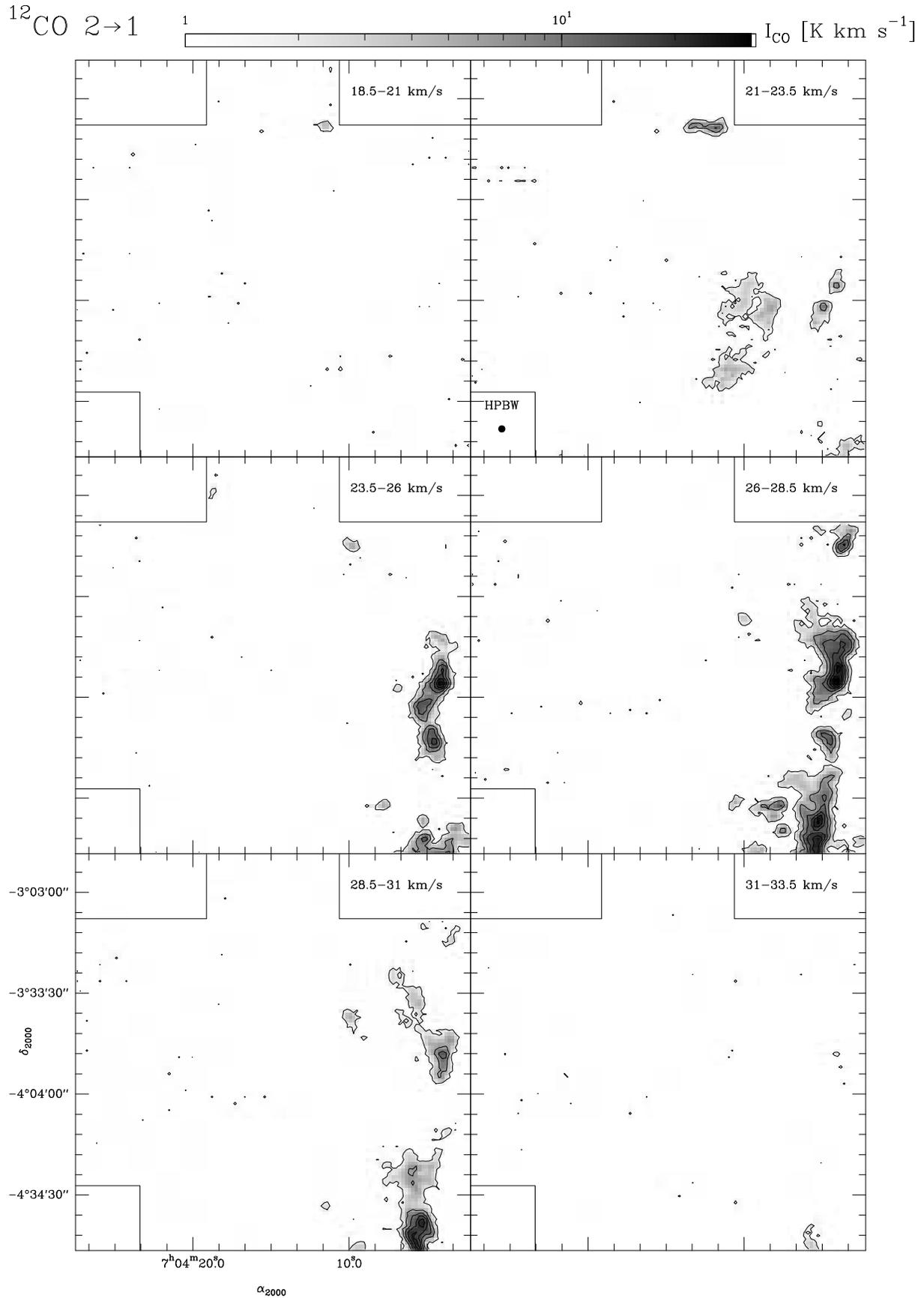}
\caption{Channel maps of $^{12}$CO (2$\to$1) integrated intensity for 
the velocity range P (18--33.5~km~s$^{-1}$) in a step of 2.5~km~s$^{-1}$. 
The integration range is given at the upper right corner of each map. 
Countours are plotted from 1.6 to 31.1~K~km~s$^{-1}$ by 4.9~K~km~s$^{-1}$ 
(5 to 95\% by 15\% of the maximum for all the six maps.)}\label{appfig1}
\end{figure*}

\begin{figure*}
\centering
\includegraphics[width=16.4cm,clip]{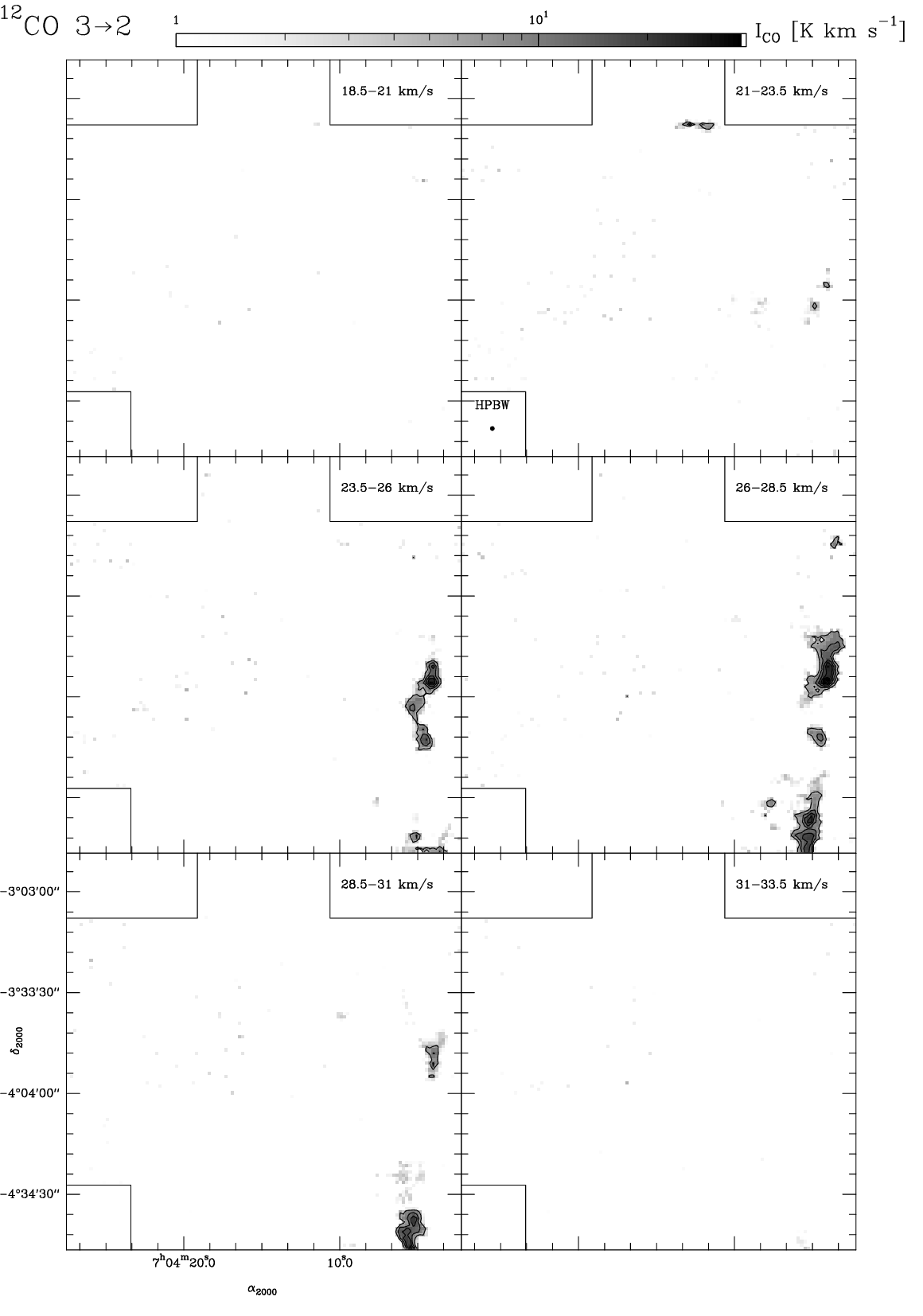}
\caption{Channel maps of $^{12}$CO (3$\to$2) integrated intensity for 
the velocity range P (18--33.5~km~s$^{-1}$) in a step of 2.5~km~s$^{-1}$. 
The integration range is given at the upper right corner of each map. 
Countours are plotted from 5.6 to 33.7~K~km~s$^{-1}$ by 5.6~K~km~s$^{-1}$ 
(15 to 95\% by 15\% of the maximum value for all the six
maps.)}\label{appfig2}
\end{figure*}

\begin{figure*}
\centering
\includegraphics[width=16.4cm,clip]{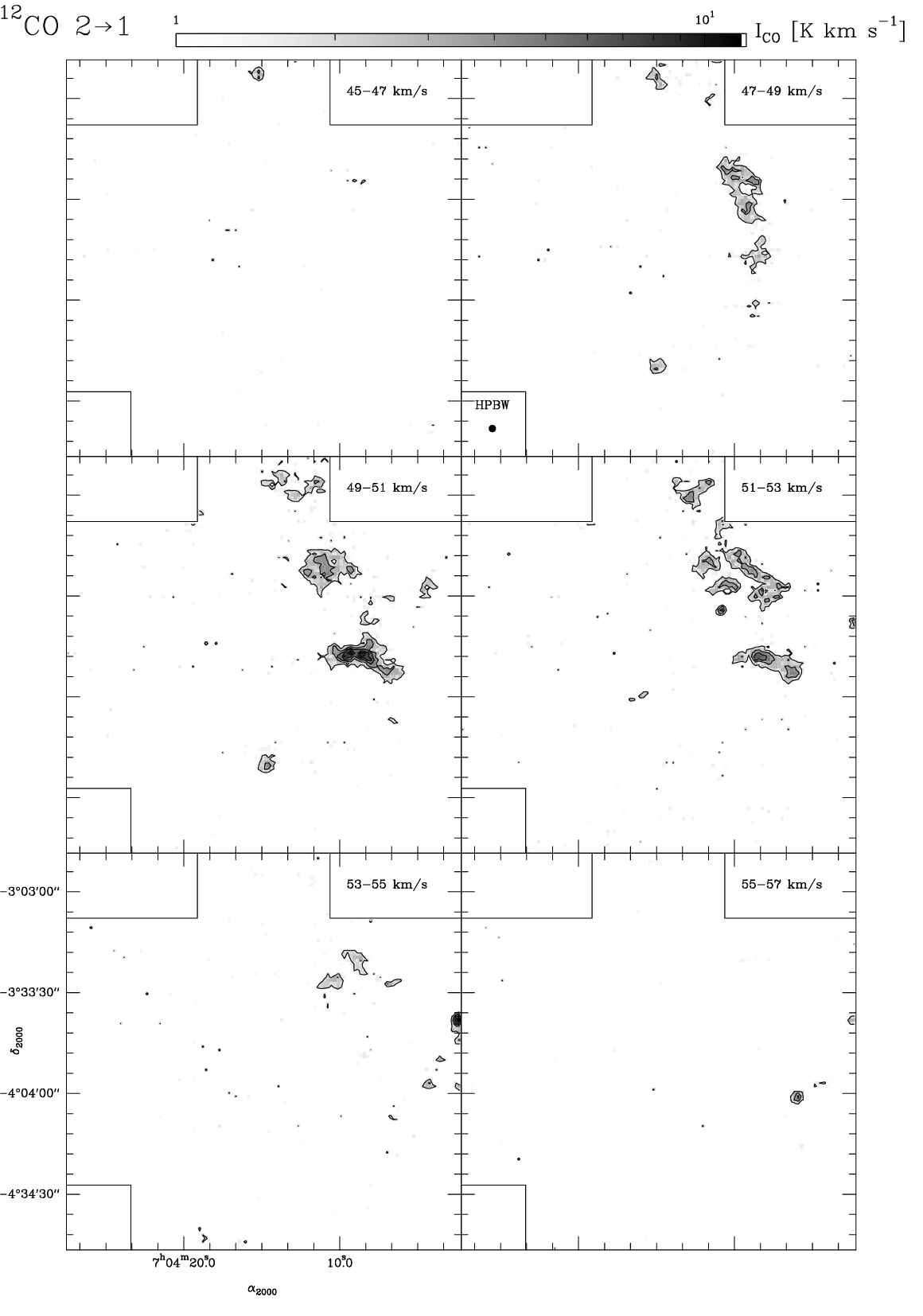}
\caption{Channel maps of $^{12}$CO (2$\to$1) integrated intensity for 
the velocity range NC (45--57~km~s$^{-1}$) in a step of 2~km~s$^{-1}$. 
The integration range is given at the upper right corner of each map. 
Countours are plotted from 1.6 to 10.6~K~km~s$^{-1}$ by 1.8~K~km~s$^{-1}$ 
(13 to 88\% by 15\% of the maximum for all the six maps.)}\label{appfig3}
\end{figure*}

\begin{figure*}
\centering
\includegraphics[width=16.4cm,clip]{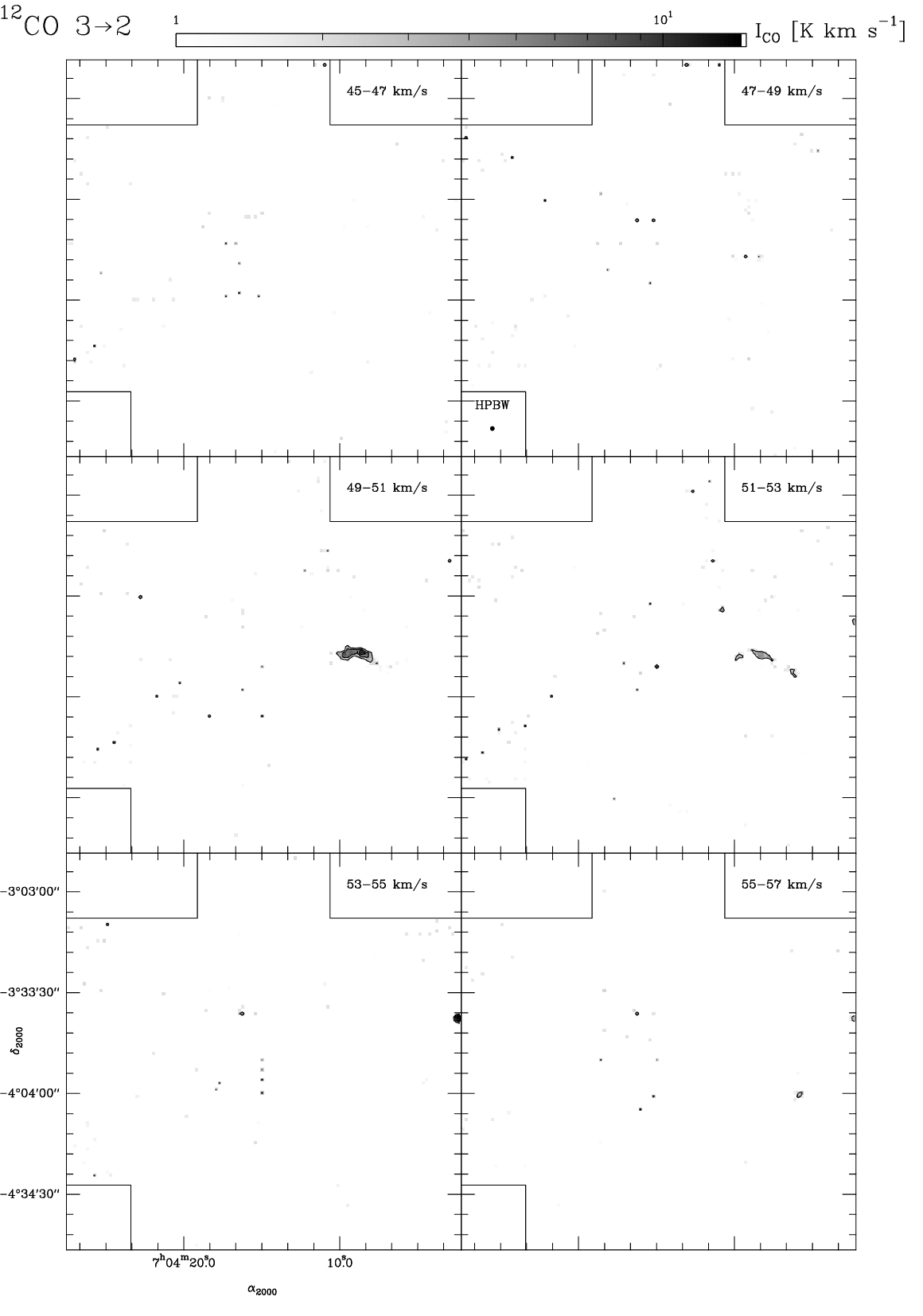}
\caption{Channel maps of $^{12}$CO (3$\to$2) integrated intensity for 
the velocity range NC (45--57~km~s$^{-1}$) in a step of 2~km~s$^{-1}$. 
The integration range is given at the upper right corner of each map. 
Countours are plotted from 2.2 to 14.2~K~km~s$^{-1}$ by 3~K~km~s$^{-1}$ 
(15 to 95\% by 20\% of the maximum for all the six maps.)}\label{appfig4}
\end{figure*}

Figures \ref{appfig1} -- \ref{appfig4} present channel maps integrated over
narrow ranges in $V_{\rm LSR}$. The data have been obtained with the KOSMA
telescope. Details of the observations can be found in Sect.~\ref{obs_sect}. 
The maps presented in Figs.~\ref{appfig1} and \ref{appfig3} have been obtained 
in the $^{12}$CO (2$\to$1) transition. The same but for the $^{12}$CO (3$\to$2)
transition is shown in Figs.~\ref{appfig2} and \ref{appfig4}.


\begin{table*}
\caption{Measured and calculated parameters for individual clouds in the
V838 Mon field. See text for the explanations of the columns.}
\label{tab_res}
\begin{tiny}
\begin{tabular}{l r r r r r r r r r r r r r r r}
\hline
\noalign{\smallskip}
nr & $l$~~~ & $b$~~~ &$\alpha_{2000}$~~~~~&$\delta_{2000}$~~~~~~& $A$~~ &
$W_{2-1}$ & $V_{2-1}$&$\Delta V_{2-1}$&$W_{3-2}$&$V_{3-2}$&$\Delta V_{3-2}$&
$d$~~& $R$~~&$M_{vir}$~&$M_X$~\\
(1) & (2)~ & (3)~ & (4)~~~~~ & (5)~~~~~~ & (6)~ & (7)~ & (8)~ & (9)~ & (10) & (11) & (12) &
(13) & (14) & (15) & (16) \\
\noalign{\smallskip}
\hline
\noalign{\smallskip}
P1   &217.07 &0.74 &06:59:08.5 &--03:16:06.2 &61  &328.3  &27.7  &2.16 &77.3   &28.0 &1.88 &3.0 &3.8  &1100  &510 \\
P2   &217.33 &1.39 &07:01:55.5 &--03:11:47.6 &67  &282.2  &21.9  &1.83 &159.3  &22.0 &1.71 &2.3 &3.1  &630   &260 \\
P3   &217.33 &1.17 &07:01:08.7 &--03:18:04.8 &22  &50.3   &24.2  &1.56 &       &     &     &2.6 &2.0  &300   &58  \\
P4   &217.37 &0.89 &07:00:12.5 &--03:28:19.1 &33  &46.8   &29.2  &1.24 &       &     &     &3.2 &3.0  &280   &83  \\
P5   &217.44 &0.74 &06:59:48.0 &--03:35:51.8 &61  &162.6  &28.8  &1.95 &24.9   &28.9 &1.28 &3.1 &4.0  &940   &280 \\
P6   &217.61 &0.52 &06:59:21.3 &--03:50:46.3 &347 &4093.0 &27.0  &3.84 &2568.1 &27.0 &3.46 &2.9 &8.8  &8100  &6000\\
P7   &217.68 &0.97 &07:01:05.7 &--03:42:30.5 &44  &108.5  &28.7  &1.46 &27.6   &28.7 &1.03 &3.1 &3.4  &440   &180 \\
P8   &217.90 &0.45 &06:59:38.5 &--04:08:11.7 &82  &732.0  &24.8  &3.42 &423.5  &24.7 &3.36 &2.6 &3.9  &2800  &890 \\
P9   &218.01 &0.84 &07:01:14.3 &--04:03:20.7 &133 &234.0  &22.7  &1.61 &19.9   &22.8 &1.45 &2.4 &4.5  &720   &230 \\
P10  &218.01 &0.70 &07:00:43.3 &--04:07:39.4 &115 &250.2  &22.1  &1.82 &55.5   &22.2 &1.61 &2.3 &4.1  &830   &240 \\
P11  &218.02 &0.35 &06:59:29.1 &--04:17:32.1 &100 &887.4  &25.5  &3.14 &533.7  &25.5 &2.85 &2.7 &4.4  &2700  &1200\\
P12  &218.32 &0.25 &06:59:43.3 &--04:36:34.6 &421 &4472.5 &28.2  &3.59 &2463.9 &28.3 &3.36 &3.1 &10.3 &8100  &7400\\
P13  &218.34 &0.74 &07:01:27.9 &--04:23:49.4 &151 &291.2  &22.0  &2.10 &16.7   &22.0 &1.70 &2.3 &4.6  &1300  &270 \\
P14  &218.45 &0.45 &07:00:38.8 &--04:37:50.0 &70  &285.2  &27.0  &2.48 &106.5  &27.0 &2.20 &2.9 &4.0  &1500  &420 \\
P15  &218.49 &0.61 &07:01:19.3 &--04:35:31.4 &17  &35.2   &27.0  &1.69 &       &     &     &2.9 &2.0  &340   &52  \\
P16  &218.51 &0.34 &07:00:23.7 &--04:44:07.2 &19  &67.9   &27.5  &1.88 &19.6   &27.3 &1.79 &2.9 &2.1  &460   &100 \\
P17  &218.56 &0.67 &07:01:39.6 &--04:37:20.1 &12  &17.1   &29.0  &1.19 &       &     &     &3.1 &1.8  &160   &30  \\
NC1  &217.17 &1.62 &07:02:25.8 &--02:57:20.3 &52  &80.1   &49.2  &4.09 &       &     &     &6.2 &7.4  &7600  &550 \\
NC2  &217.18 &1.73 &07:02:51.7 &--02:54:43.2 &31  &56.9   &47.1  &3.40 &       &     &     &5.9 &5.3  &3800  &340 \\
NC3  &217.19 &1.47 &07:01:56.4 &--03:02:34.5 &100 &191.4  &50.9  &3.19 &23.0   &50.6 &5.43 &6.6 &10.8 &6800  &1500\\
NC4  &217.32 &0.68 &06:59:23.2 &--03:31:12.2 &25  &27.1   &49.9  &1.77 &       &     &     &6.4 &5.2  &1000  &200 \\
NC5  &217.45 &0.90 &07:00:24.0 &--03:32:25.6 &69  &147.8  &52.0  &2.66 &14.2   &51.2 &2.97 &6.8 &9.2  &4000  &1200\\
NC6  &217.46 &1.15 &07:01:18.6 &--03:25:47.6 &258 &850.4  &50.7  &5.33 &69.3   &50.1 &4.01 &6.5 &17.2 &30000 &6400\\
NC7  &217.57 &0.96 &07:00:51.3 &--03:36:47.4 &78  &36.0   &48.1  &1.72 &27.2   &47.9 &1.38 &6.0 &8.8  &1600  &230 \\
NC8  &217.65 &1.10 &07:01:29.1 &--03:37:29.3 &15  &30.7   &52.4  &1.95 &       &     &     &6.9 &4.4  &1000  &260 \\
NC9  &217.76 &0.45 &06:59:21.6 &--04:01:03.3 &24  &33.6   &54.7  &1.61 &       &     &     &7.4 &5.9  &950   &320 \\
NC10 &217.77 &0.80 &07:00:38.6 &--03:51:37.7 &250 &900.9  &50.3  &3.08 &285.0  &50.3 &2.63 &6.5 &16.7 &9800  &6600\\
NC11 &217.90 &0.54 &06:59:57.3 &--04:06:07.0 &20  &42.4   &55.5  &2.92 &13.6   &56.1 &1.57 &7.5 &5.5  &2900  &420 \\
NC12 &218.48 &1.01 &07:02:43.1 &--04:23:55.3 &47  &83.5   &49.2  &2.85 &9.6    &49.3 &2.49 &6.3 &7.0  &3500  &580 \\
\noalign{\smallskip}
\hline
\end{tabular}
\end{tiny}
\end{table*}

As can be seen from Figs. \ref{appfig1} -- \ref{appfig4} numerous more or less
separate clouds of the CO emission can be distinguished. We have measured
principal observational parameters of the clouds and the results are given
in Table~\ref{tab_res}. Column (1) gives the cloud number. The
galactic ($l$ and $b$) and equatorial ($\alpha_{2000}$ and $\delta_{2000}$) 
coordinates of the geometric cloud centre are displayed in columns (2)--(5). 
The surface area of the cloud (in squared arcmin),
determined from the $J=2$$\to$$1$ map, can be found
in column (6). Columns (7), (8) and (9) give the total intensity
integrated over the cloud surface  
(in ${\rm K}\,{\rm km}\,{\rm s}^{-1}$), the mean value
of the cloud $V_{\rm LSR}$ and the FWHM of the velocity profile 
(both in ${\rm km}\,{\rm s}^{-1}$), as measured in 
the $J=2$$\to$$1$ transition. 
The latter two values
have been obtained from fitting a Gaussian profile to the $V_{\rm LSR}$ profile
integrated over the cloud surface. The same three measurements, but obtained
from the $J=3$$\to$$2$ map, are shown in columns (10), (11) and (12).

The upper part of Table~\ref{tab_res} lists clouds
observed in the $V_{\rm LSR}$ range $18-32$~km~s$^{-1}$, 
called range P, later on. Those seen in the
range $44-57$~km~s$^{-1}$, in the following
called range NC, are given in the bottom part of the table. 
Note that in certain cases 
definition of an individual cloud is not obvious and can be subjective. 
Our goal was to delimit regions physically related.
Therefore we have
analysed not only the surface brightness distribution but also the
surface distribution of $V_{\rm LSR}$ and attempted to single out
structures concise in the space of both parameters. Small regions at the
edge of our survey have not been
included in Table~\ref{tab_res} as, probably, they are parts
of larger structures lying outside the survey.

The data in Table \ref{tab_res} allow us to estimate several important
physical parameters of the CO clouds. The radial velocities can be used to 
estimate kinematic distances to the objects. For this purpose we have used
the mean values of $V_{\rm LSR}$ derived from the 
$J=2$$\to$$1$ survey and listed in column (8). 
Note that the velocities derived from the
$J=3$$\to$$2$ survey (column 11) do not significantly differ from the
above ones. We have applied the Galactic
rotational curve of Brandt \& Blitz (\cite{bb93}) and the resultant
heliocentric distances (in kpc) are given in column (13) in
Table~\ref{tab_res}. As can be seen from the results, the 
clouds in the $V_{\rm LSR}$ range P are typically at a distance of
$2.3-3.2$~kpc and form a different population from those in the range
NC, which are at $6.0-7.5$~kpc. These two populations can be easily
indentified with two outer Galactic arms, i.e. the Perseus arm and the
Norma-Cygnus arm (Russeil \cite{russ03}), respectively. (Hence our
notation in the cloud numbers, P or NC, in Table~\ref{tab_res}).

Assuming the virial theorem the mass of a molecular cloud can be estimated
from (e.g. MacLaren et~al. \cite{mcl88})
\begin{equation}
  M_{\rm vir} = k_2\,R\,\Delta V^2,
\label{vir_eq}
\end{equation}
where $M_{\rm vir}$ is the cloud mass in $M_\odot$, $R$ is the cloud radius
in parsecs and $\Delta V$ is the FWHM of the integrated line profile in
${\rm km}\,{\rm s}^{-1}$. $k_2$ is a constant whose value depends on the
density distribution in the cloud. Following a discussion of
observational results for clouds in the outer Galaxy done in MacLaren
et~al. (\cite{mcl88}) we assume an $r^{-2}$ density distribution and 
$k_2 = 126$. (Note that $k_2$ rather weakly depends on the density
distribution, e.g. $k_2 = 190$ for $\rho \sim r^{-1}$.)

The cloud radius, as given in column~(14) in Table~\ref{tab_res} (in pc), 
has been calculated from $R = \sqrt{S/\pi}$, where $S$ is
the cloud surface obtained form the measured angular surface given in
column (6) and the distance in column (13) in Table~\ref{tab_res}.

As can be seen from Table~\ref{tab_res},
the velocity dispersion, $\Delta V$,
in the $J=3$$\to$$2$ transition (column 12) is systematically lower than
that in the $J=2$$\to$$1$ transition (column 9). Excluding the cloud NC3, for
which the $J=3$$\to$$2$ profile is considerably affected by noise, the mean value
and the standard deviation of $\Delta V(3-2)/\Delta V(2-1)$ is
$0.86\pm0.12$. Since the $J=3$$\to$$2$ transition is expected to be less
optically thick than the $J=2$$\to$$1$ one, the above effect
suggests that a certain line broadening due to optical thickness
effects can be present. Therefore, following
results of MacLaren et~al. (\cite{mcl88}), we have reduced the FWHM value
from column~(9) in Table~\ref{tab_res} by a factor of 0.7 when substituting to
Eq.~(\ref{vir_eq}). The resultant cloud masses, $M_{\rm vir}$,
are given in column~(12) in Table~\ref{tab_res} (in $M_\odot$).

Another method of estimating the cloud mass is based on a conversion
factor, $X_{\rm CO}$, which allows one to derive the column density of 
${\rm H}_2$ molecules along the line of sight from the observed CO
velocity-integrated line intensity. The resultant mass, $M_{\rm X}$,
can then be obtained from   
\begin{equation}
  M_{\rm X} = 1.4\ m_{{\rm H}_2}\ X_{\rm CO}\ W_{\rm CO}\ d^2,
\label{xco_eq}
\end{equation}
where $W_{\rm CO}$ is the line intensity integrated over the velocity and
the cloud surface, while $d$ is the distance. $m_{{\rm H}_2}$ is the mass of an
${\rm H}_2$ molecule and the factor of 1.4 corrects for helium. Note that
$X_{\rm CO}$ is usually meant to convert intensities in the CO~$J =
1$$\to$0 line, whereas our measurements were done in the $J = 2$$\to$$1$
line. However, as observations of molecular clouds show, the ratio of the two
lines is usually close to 1.0 (e.g. Brand \&
Wouterloot \cite{BW95}, Hasegawa \cite{hase97}). Therefore we can 
directly use our measurements of $W_{\rm CO}$ given in column~(7) in
Table~\ref{tab_res} when calculating the cloud masses from Eq.~(\ref{xco_eq}).

There are numerous estimates of the value of $X_{\rm CO}$ in the
literature. For the inner Galaxy they are close to 1.0 (in units 
$10^{20}\,{\rm molecules}\ {\rm cm}^{-2}\,{\rm K}^{-1}\,
{\rm  km}^{-1}\,{\rm s}$)  (e.g. MacLaren et~al. \cite{mcl88}, Maddalena
et~al. \cite{mad86}). In the outer Galaxy $X_{\rm CO}$ is
usually larger (Mead \& Kutner \cite{mk88}, Digel et~al. \cite{dig90},
Sodroski \cite{sod91}, Usuda
et~al. \cite{usuda99}). We assume, as a working value, $X_{\rm CO} = 1.0$. 
The resultant cloud masses (in $M_\odot$) obtained from Eq.~(\ref{xco_eq}) 
and using the values from columns (7) and (13) in Table~\ref{tab_res} can be 
found in the last column of the table.

\begin{figure}
 \resizebox{\hsize}{!}{\includegraphics{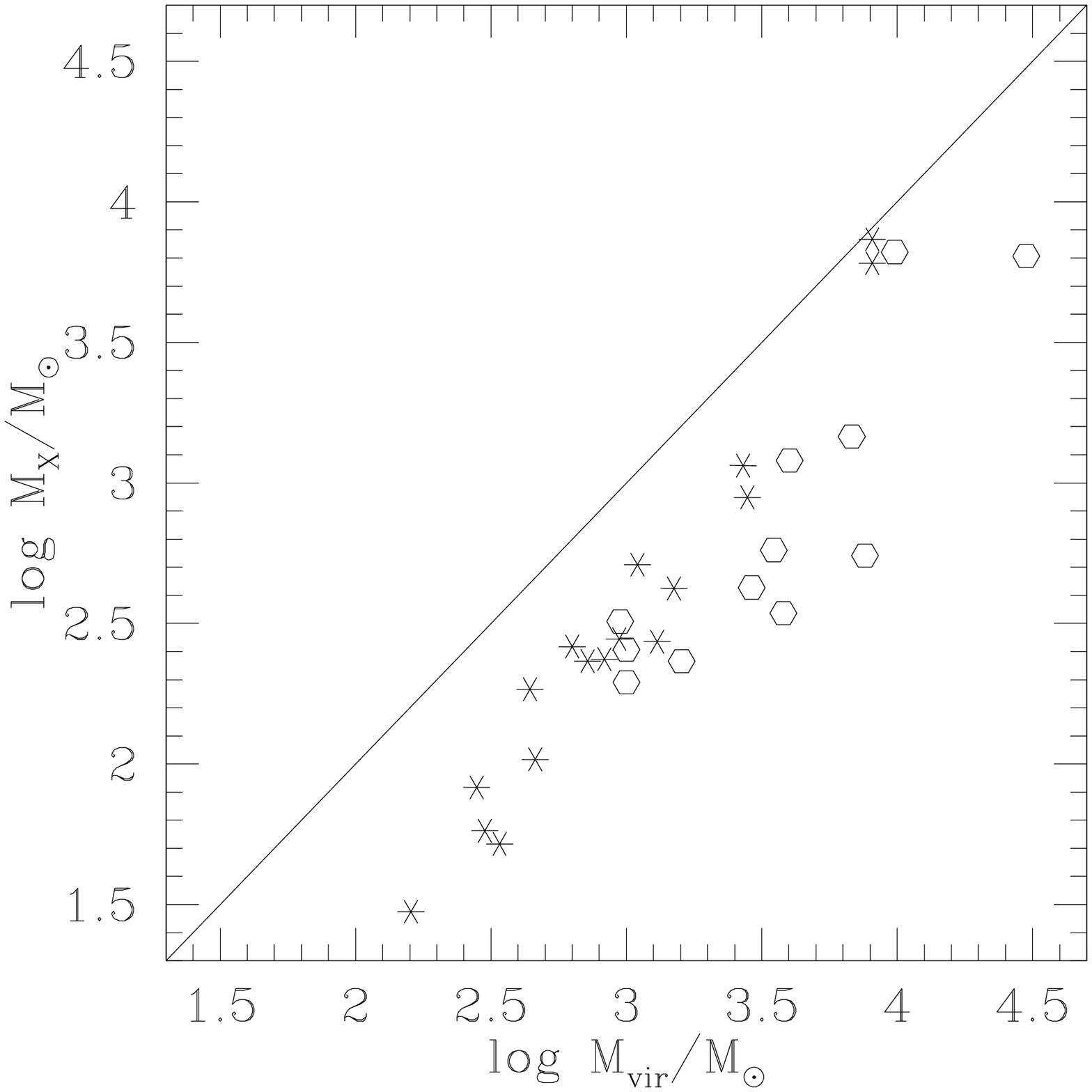}}
 \caption{$M_{\rm X}$ (derived from Eq.~\ref{xco_eq}, see column~16 in
Table~\ref{tab_res}), plotted versus $M_{\rm vir}$ 
(derived from Eq.~\ref{vir_eq}, see column~15 in
Table~\ref{tab_res}).
Asterisks: P clouds, open symbols: NC clouds. The
full line shows a 1:1 relation.}
 \label{mx_mv_fig}
\end{figure}

Fig. \ref{mx_mv_fig} compares the resultant clouds masses determined from
the above two methods. As can be seen, for all the clouds $M_{\rm vir}$ is
greater than $M_{\rm X}$. This is in accord with the above note that in the
outer Galaxy $X_{\rm CO}$ is expected to be greater than 1. The clouds
from the Perseus arm, however, suggest a linear relation in
Fig.~\ref{mx_mv_fig} (see asterisks) with a slope greater than 1. The observed
position in the figure can be fitted by 
${\rm log}\,M_{\rm X} = -1.67 + 1.40\ {\rm log}\,M_{\rm vir}$, which can be
transformed to $X_{\rm CO} = M_{\rm vir}/M_{\rm X}
= 2.2 (M_{\rm X}/10^3 M_\odot)^{-0.29}$. Note, however, that the statistics
is poor and the above relation relies on a few extreme points in
Fig.~\ref{mx_mv_fig}. No similar relation is seen for the Norma-Cygnus arm
clouds (open points in Fig.~\ref{mx_mv_fig}). A mean value of the shift of
the open points from the full line in Fig.~\ref{mx_mv_fig} corresponds to 
$X_{\rm CO} = 5.1$. This result is close to the value found for the outer
Galaxy by Usuda et~al. \cite{usuda99}, i.e. $X_{\rm CO} =
5.4\pm0.5$. In the NC population we do not see clouds of a mass as 
low as some of the P clouds. If a similar mass range is considered in both
populations, i.e. when the 5 least massive P clouds are not taken into
account, then a mean value of $X_{\rm CO} = 2.6$ is obtained for the Perseus
arm clouds. 
\end{appendix}

\end{document}